\documentclass{pic2012}

\usepackage{graphicx}
\usepackage{xspace}
\usepackage{amsmath}

\def\runauthor{Maeda Yosuke}
\def\shorttitle{Charged-particle veto detector in  \KOTO experiment}

\newcommand{\KOTO}{K$^{\mathrm{O}}$TO\xspace}
\newcommand{\kpnn}{$K_{L} \rightarrow \pi^{0} \nu \bar{\nu}$\xspace}
\newcommand{\Kpnn}{$K^{+} \rightarrow \pi^{+} \nu \bar{\nu}$\xspace}
\newcommand{\kpp}{$K_{L} \rightarrow \pi^{0} \pi^{0}$\xspace}
\newcommand{\kep}{$K_{L} \rightarrow e \pi \nu$\xspace}
\newcommand{\kmp}{$K_{L} \rightarrow \mu \pi \nu$\xspace}
\newcommand{\p}{$\pi^{0}$\xspace}
\newcommand{\KL}{$K_{\mathrm{L}}$\xspace}
\newcommand{\g}{$\gamma$\xspace}

\begin{document}

\title{Charged-particle veto detector\\
for the \kpnn study in the J-PARC \KOTO experiment}

\author{Yosuke Maeda}

\address{Department of Physics, Kyoto University\\
Kitashirakawa-Oiwakecho, Sakyo-ku, Kyoto, 606-8502, Japan\\
E-mail: maeda\_y@scphys.kyoto-u.ac.jp }

\maketitle

\abstracts{
A charged-particle veto detector was constructed to study the rare decay \kpnn
in the J-PARC \KOTO experiment.
This detector consists of 3mm-thick plastic scintillator strips
and wavelength shifting fibers coupled with MPPCs at the both ends,
which makes it possible to obtain large light output over the wide region.
After the construction and installation to the \KOTO experimental area,
its performance was studied using charged particles from \KL decays.
As a preliminary result,
we found that the detector had the light yield larger than 10p.e./100keV
and the timing resolution better than 3ns for most regions,
which satisfied the requirements to achieve the standard model sensitivety($\sim10^{-11}$)
for the \kpnn detection.
}

\section{\KOTO experiment for \kpnn}
The decay \kpnn is a CP-violating flavor changing neutral current (FCNC) process
and is strongly suppressed in the standard model (SM).
Its branching ratio is predicted to be $(2.43\pm0.39)\times10^{-11}$
\cite{SMPrediction} in the SM,
where the theoretical uncertainty is reduced to be small.
Since the contribution of new particles to the loop diagram
can change the decay rate drastically,
this process is an excellent probe for the physics beyond the SM.

%The Fig.\ref{fig:NewPhysics} shows predicted \kpnn branching ratio
%for various new physics models with its charged mode, \Kpnn.

Due to the isospin symmetry, an upper limit for the branching ratio Br(\kpnn) can be set
from the \Kpnn experimental limit.
By using the latest result from the BNL E787 and E949 experiments\cite{E949},
we obtain
\begin{equation}
  \mathrm{Br}(K_{L} \rightarrow \pi^{0} \nu \bar{\nu})<1.4\times10^{-9}.
\end{equation}
This indirect limit is called "Grossman-Nir bound"\cite{GNlimit},
and an experiment with a higher sensitivity than this means
to search physics beyond the SM.
On the other hand, the current experimental limit by the direct search
is given as $2.6\times10^{-8} (90\% \mathrm{C.L.})$
by KEK E391a experiment\cite{E391a},
which is higher than the above limit by one order of magnitude.

%\begin{figure}[!thb]
%\label{fig:NewPhysics}
%\vspace*{7.0cm}
%\begin{center}
%\special{psfile=figure/NewPhysics.png voffset=-60 vscale=40 hscale= 40 hoffset=10 angle=0}
%\includegraphics[height=5cm1qwcv ]{figure/Krare_BSM_new.eps}
%\centerline{\epsfxsize=2.9in\epsfbox{kim_mephi_lep.ps}}
%\caption[*]{
%Prediction of branching ratio for \kpnn and \Kpnn decay
%by various physics models beyond the Standard Model.
%}
%\end{center}
%\end{figure}

The \KOTO experiment is the successor of E391a to search for this decay
using an intense proton beam of Japan Accelerator Research Complex(J-PARC).
The same experimental techniques are used with lots of improvements.
The ultimate goal is to achieve the SM sensitivity,
and the first physics data taking will be performed in 2013,
where a better sensitivity than the Grossman-Nir bound is expected.

\section{signal and background}
The \KOTO experiment is performed in the Hadron Experimental Facility of J-PARC
using the 30GeV slowly-extracted proton beam.
The primary proton beam hits the production target
(nickel, platinum or gold; according to the beam power)
and the narrow neutral beam is obtained from the secondary particles
with the beamline consisting of collimators and a sweeping magnet.
At the downstream of the beamline, the detector is placed
to observe the \KL decay from the neutral beam.

Since it is not possible to tag the incident \KL and detect outgoing neutrinos,
a single \p is the only visible particle in this decay process.
Thus, we identify \kpnn events where only an isolated \p exists;
two photos from \p are detected by an electromagnetic calorimeter
and no visible energy in the "veto detectors"
which surround the whole decay region.

There are two types of background  :  the \KL-induced one and the neutron-induced one.
The former is due to detector inefficiency.
When extra particles from \KL are not detected,
such events can be regarded as signals.
Since \kpp decay has a similar topology to the signal event
and a small number of extra particle (two photons from the additional \p),
this is the most serious background in the \KOTO experiment.
The Ke3(\kep) and K$\mu$3(\kmp) decays,
in which two oppositely charged particles are emitted,
have a large branching ratio
and can fake as signals when both charged particles are not identified.

The latter one is due to the \p production by neutrons in the beam halo.
When they hit materials around beam halo such as detector itself,
\p or $\eta$ can be generated.
These events have two photons in the final state when no other particles produced or detected,
and make backgrounds.

\section{charged-veto detector}
The \KOTO electromagnetic calorimeter, whose diameter is 2m,
consists of 2,716 undoped CsI crystals.
The whole surface is covered by "Charged Veto" detector (CV),
and a cluster of hit crystals without any output in CV is identified as a \g.
When the CV fails to detect charge particles entering the calorimeter,
they are misidentified as a photon, and can make background.
Thus the inefficiency for charged particles must be suppressed to a low level for the wide region.
On the other hand,
since the detector itself can be a source of \p production
by neutrons in the beam halo,
the amount of materials should also be small.

To satisfy both requirements,
we use 3mm-thick plastic scintillators (Saint-Gobain BC404)
with wavelength shifting fiber (Kuraray Y11).
Multi Pixel Photon Counters (MPPCs) by Hamamatsu are coupled
to the both ends of fibers.
The long attenuation length of the wavelength shifting fibers
and high photon detection efficiency of MPPC
enables us to get large light yield over the wide range,
while the amount of materials is kept as small as possible.

The detector consists of two planes separated by 25cm each other.
The front(rear) plane contains 48(44) modules,
each of which is the combination of a scintillator strip, wavelength shifting fibers and MPPCs.
The layout of the scintillator strips and the design of a module are shown in Fig.1.
%\ref{fig:PlaneConfig}.
There are 7 U-shaped grooves to set fibers with 1cm interval in a strip,
and wavelength shifting fibers are set in each groove.
Both sides of the strips where the neighboring strips come have
an angle of 30$^{\circ}$ to the beam direction to eliminate the gaps between strips.

\begin{figure}[!thb]
\label{fig:PlaneConfig}
%\vspace*{7.0cm}
\begin{center}
%\special{psfile=figure/PlaneConfig.png voffset=-60 vscale=40 hscale= 40 hoffset=10 angle=0}
\includegraphics[width=120mm]{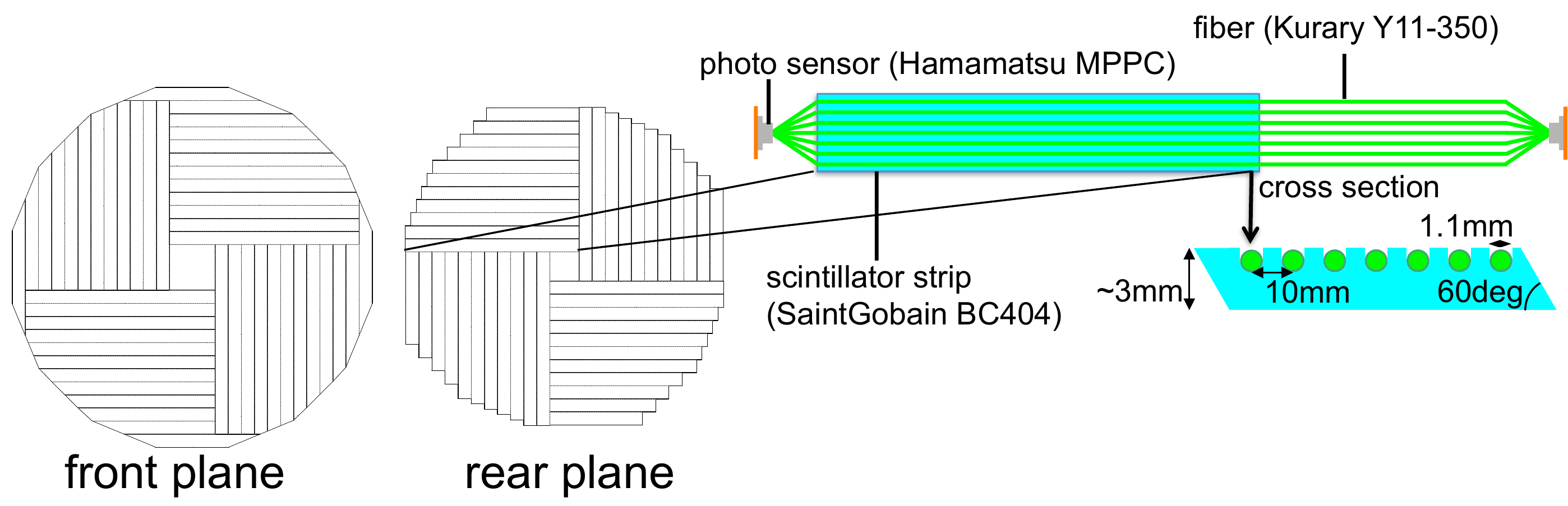}
%\centerline{\epsfxsize=2.9in\epsfbox{kim_mephi_lep.ps}}
\caption[*]{
Layouts of modules in each plane(left)
;design of a module(right).
}
\end{center}
\end{figure}

\section{module production and construction}
The scintillator strips for CV was cut from cast sheet with the size of 60cm $\times$ 100cm.
The effect due to the non-uniformity of the scintillator thickness
was studied by a Monte Carlo simulation,
and the thickness of all sheets were measured.
The path length of a particle in a scintillator is
so short that the inefficiency due to photon statistics can increase drastically in a region
where the grooves is too deep or thickness of a scintillator is too thin.
As a result of a simulation,
we found that the minimum thickness under grooves should be larger than 1.8mm
to reduce the inefficiency less than $10^{-3}$.
The thickness for all cast sheet was measured using a laser displacement sensor.
The maps of the measured thickness for each sheet was used
to determine the thickness under grooves and the module assignment.
Here regions thinner than 2.75mm was avoided in order to keep the thickness under grooves
and not to make the grooves too shallow.

For the effective fabrication of the modules and the uniformity of quality among modules,
an automatic gluing system shown in Fig.2 %\ref{fig:GluingSystem}
was developed
to glue the grooved scintillator and the wavelength shifting fibers.
Optical cement(EJ500) with lowered viscosity was used,
where the mixture ratio is changed to resin:hardener=3.6:1.4 in weight
from 4:1 in nominal to fit the system.
The quantity of glue to apply was controlled by adjusting the stage speed.
After the fabrication, the position dependence of light yield was inspected for each module
by using a $^{90}$Sr $\beta$ ray source.
It was confirmed that the enough light yield was achieved
for all the modules in all the position.

These modules were shipped to J-PARC
and the assembly was done.
The support structure consisted of an aluminum frame with hexadecagon shape
and the carbon fiber reinforced plastic(CFRP) plate.
Each module was mounted on the CRRP plate
and the readout electronics is equipped on the aluminum frame.
Fig.3 %\ref{fig:assembly}
is the photo of the construction.

\begin{figure}[htbp]
 \begin{minipage}{0.49\hsize}
  \begin{center}
   \includegraphics[width=50mm]{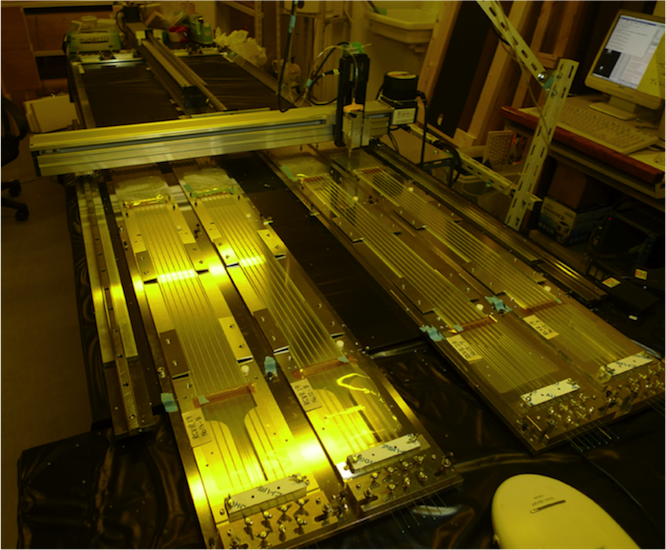}
   \label{fig:GluingSystem}
  \end{center}
  \caption{
Photo of the fiber gluing system.
A syringe containing the glue was mounted on electrical stages
and its ejection was controlled by the dispenser
synchronized with the stages.
4 modules with identical shape can be glued at the same time.
  }
  \label{fig:one}
 \end{minipage}
 \begin{minipage}{0.49\hsize}
  \begin{center}
   \includegraphics[width=50mm]{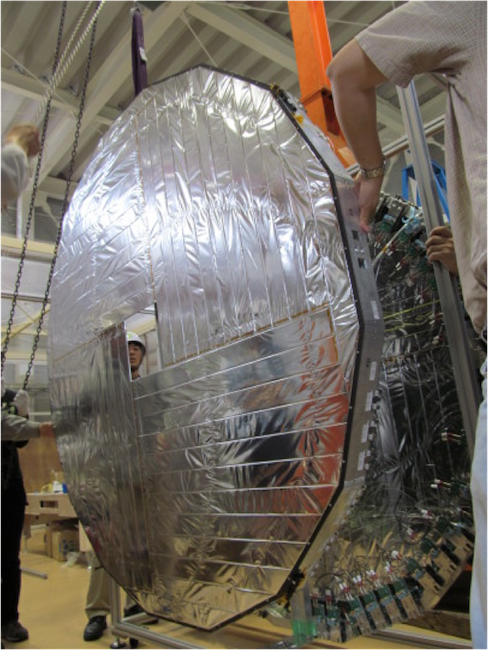}
   \label{fig:assembly}
  \end{center}
  \caption{Photo in combining the two planes after the assembly of each plane.}
 \end{minipage}
\end{figure}

\section{performance test with \KOTO neutral beam}
The CV detector was installed to the \KOTO experimental area
and its performance was tested using two charged particles from \KL decay.
A series of drift chambers were placed in front of CV,
and the events with two charged particle are triggered
by the scintillator hodoscope located in the downstream of CV.
Signals from MPPCs were read by the same 125MHz FADC modules with Bessel filters
which are used in the readout of the CsI calorimeter.

The one photoelectron gain of all MPPCs were monitored
by the calibration system with blue LEDs.
Fig.4 %\ref{fig:GainCalibration}
shows an example of waveform recorded by FADC.
The sum of 25 samples around the signal timing was used to evaluate the output for each channel,
and the output corresponding to the one-photoelectron pulse
was stable within $\pm3\%$ over 5 days.

\begin{figure}[!thb]
\label{fig:GainCalibration}
%\vspace*{7.0cm}
\begin{center}
%\special{psfile=figure/PlaneConfig.png voffset=-60 vscale=40 hscale= 40 hoffset=10 angle=0}
\includegraphics[width=75mm]{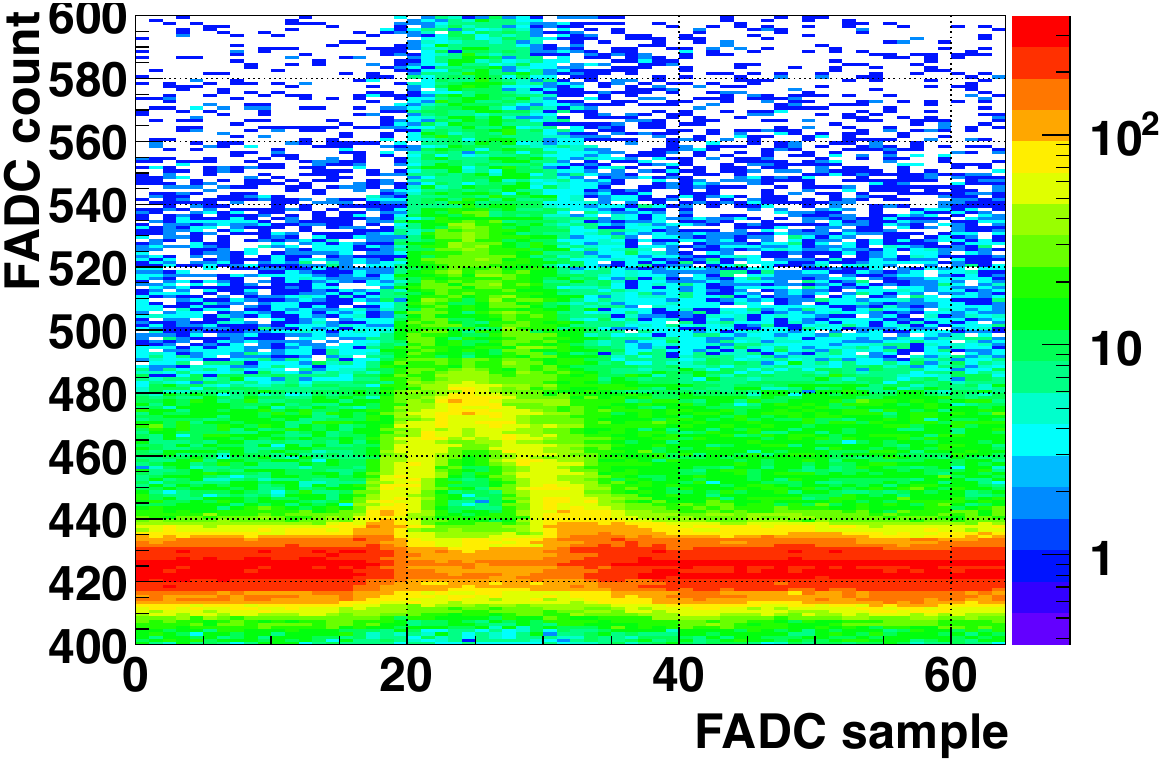}
%\centerline{\epsfxsize=2.9in\epsfbox{kim_mephi_lep.ps}}
\caption[*]{
An example of waveform data for one photoelectron calibration run using LED.
64 samples were recorded for each event
and LED trigger events are overlaid as a two-dimension histogram in this figure.
Pedestal events and one photoelectron events are clearly seperated.
}
\end{center}
\end{figure}

The position dependence of light yield and timing resolution was evaluated for each module
by using the hit position of the charged particles obtained by drift chambers.
The light yield was defined as the integral, calculated with the same method described above,
normalized by the one photoelectron output,
and the timing resolution was estimated as the $\sigma$ of gaussian fit
for the distribution of the hit timing difference between two hits in the same module.
The result is shown in Fig.5. %\ref{fig:result}.
For the most region, a higher light yield than 60 photoelectron
and a timing resolution better than 3ns was achieved,
which satisfied the requirement of the \KOTO experiment
to suppress the background the accidental loss enough.

\begin{figure}[!thb]
\label{fig:result}
%\vspace*{7.0cm}
\begin{center}
%\special{psfile=figure/PlaneConfig.png voffset=-60 vscale=40 hscale= 40 hoffset=10 angle=0}
\includegraphics[width=120mm]{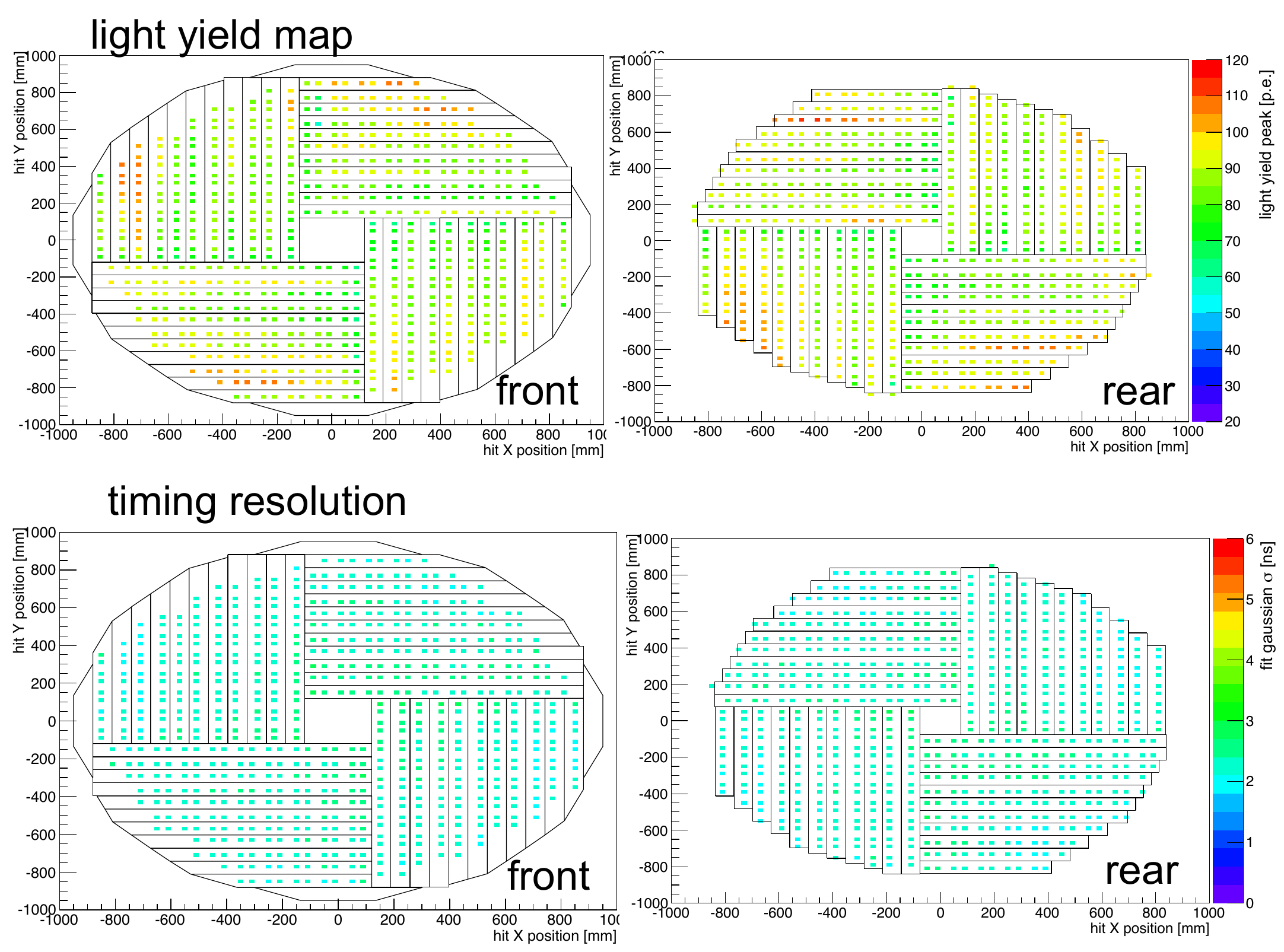}
%\centerline{\epsfxsize=2.9in\epsfbox{kim_mephi_lep.ps}}
\caption[*]{
Measured position dependence of the light yield(top)
and the timing resolution(bottom) for each plane.
}
\end{center}
\end{figure}

\section{summary \& prospect}
A charged-paritcle veto detector was successfully constructed
to study the CP-violating rare decay \kpnn in J-PARC \KOTO experiment.
Its performance was tested using \KL decay particle in the neutral \KL beam line.
It was confirmed that the detector has an enough light yield and a goal timing resolution
to active the SM sensitivity.
The construction of whole \KOTO detector is in progress
and the first physics run is schedules in 2013.

\end{document}